\documentclass[onecolumn,3p]{elsarticle}

\usepackage{graphicx}
\usepackage{amssymb}
\usepackage{amsmath}
\usepackage{hyperref}

\journal{Computer Physics Communications}

\begin{document}

\begin{frontmatter}

\title{CUDA programs for solving the time-dependent dipolar Gross-Pitaevskii equation in an anisotropic trap}

\author[scl]{Vladimir Lon\v{c}ar\corref{author}}
\ead{vladimir.loncar@ipb.ac.rs}

\author[scl]{Antun Bala\v{z}}
\ead{antun.balaz@ipb.ac.rs}

\author[scl]{Aleksandar Bogojevi\'{c}}
\ead{aleksandar.bogojevic@ipb.ac.rs}

\author[dmi]{Srdjan \v{S}krbi\'{c}}
\ead{srdjan.skrbic@dmi.uns.ac.rs}

\author[bdu]{Paulsamy Muruganandam}
\ead{anand@cnld.bdu.ac.in}

\author[ift]{Sadhan K. Adhikari}
\ead{adhikari@ift.unesp.br}

\cortext[author] {Corresponding author.}
\address[scl]{Scientific Computing Laboratory, Institute of Physics Belgrade, University of Belgrade, Pregrevica 118, 11080 Belgrade, Serbia}
\address[dmi]{Department of Mathematics and Informatics, Faculty of Sciences, University of Novi Sad, Trg Dositeja Obradovi\' ca 4, 21000 Novi Sad, Serbia}
\address[bdu]{School of Physics, Bharathidasan University, Palkalaiperur Campus, Tiruchirappalli -- 620024, Tamil Nadu, India}
\address[ift]{Instituto de F\'{\i}sica Te\'{o}rica, UNESP -- Universidade Estadual Paulista,  01.140-70 S\~{a}o Paulo, S\~{a}o Paulo, Brazil}

\begin{abstract}
In this paper we present new versions of previously published numerical programs for solving the dipolar Gross-Pitaevskii (GP) equation including the contact interaction in two and three spatial dimensions in imaginary and in real time, yielding both stationary and non-stationary solutions. New versions of programs were developed using CUDA toolkit and can make use of Nvidia GPU devices. The algorithm used is the same split-step semi-implicit Crank-Nicolson method as in the previous version (R. Kishor Kumar et al. (2015)) \cite{dbec2015}, which is here implemented as a series of CUDA kernels that compute the solution on the GPU. In addition, the Fast Fourier Transform (FFT) library used in the previous version is replaced by cuFFT library, which works on CUDA-enabled GPUs. We present speedup test results obtained using new versions of programs and demonstrate an average speedup of 12 to 25, depending on the program and input size.
\end{abstract}

\begin{keyword}
Bose-Einstein condensate; Dipolar atoms; Gross-Pitaevskii equation; Split-step Crank-Nicolson scheme; Real- and imaginary-time propagation; C program; GPU; CUDA program; Partial differential equation

\PACS 02.60.Lj; 02.60.Jh; 02.60.Cb; 03.75.-b
\end{keyword}

\end{frontmatter}

\begin{small}
\noindent
{\bf New version program summary}

\noindent
{\em Program title:} DBEC-GP-CUDA package, consisting of: (i) imag2dXY-cuda, (ii) imag2dXZ-cuda, (iii) imag3d-cuda, (iv) real2dXY-cuda, (v) real2dXZ-cuda, (vi) real3d-cuda.\\
{\em Catalogue identifier:} AEWL\_v2\_0 \\
{\em Program Summary URL:} \href{http://cpc.cs.qub.ac.uk/summaries/AEWL_v2_0.html}{http://cpc.cs.qub.ac.uk/summaries/AEWL\_v2\_0.html}\\
{\em Program obtainable from:} CPC Program Library, Queen's University of Belfast, N. Ireland.\\
{\em Licensing provisions:} Standard CPC licence, http://cpc.cs.qub.ac.uk/licence/licence.html.\\
{\em No. of lines in distributed program, including test data, etc.:} 18297.\\
{\em No. of bytes in distributed program, including test data, etc.:} 128586.\\
{\em Distribution format:} tar.gz.\\
{\em Programming language:} CUDA C.\\
{\em Computer:} Any modern computer with Nvidia GPU with Compute Capability 2.0 or higher, with CUDA toolkit (compiler and runtime, with cuFFT library, minimum version 6.0) installed. \\
{\em Operating system:} Linux.\\
{\em RAM:} With provided example inputs, programs should run on a computer with 512 MB GPU RAM. There is no upper limit to amount of memory that can be used, as larger grid sizes require more memory, which scales as NX*NY or NX*NZ (in 2d) or NX*NY*NZ (in 3d). All programs require roughly the same amount of CPU and GPU RAM. \\
{\em Number of processors used:} One CPU core and one Nvidia GPU. \\
{\em Classification:} 2.9, 4.3, 4.12.\\
{\em External routines/libraries:} CUDA toolkit, version 6.0 or higher, with cuFFT library. \\
{\em Catalogue identifier of previous version:} AEWL\_v1\_0.\\
{\em Journal reference of previous version:} Comput. Phys. Commun. 195 (2015) 117.\\
{\em Does the new version supersede the previous version?:} No. \\

\noindent\\
{\em Nature of problem:}
These programs are designed to solve the time-dependent nonlinear partial differential Gross-Pitaevskii (GP) equation with contact and dipolar interactions in two or three spatial dimensions in a harmonic anisotropic trap. The GP equation describes the properties of a dilute trapped Bose-Einstein condensate.

\noindent\\
{\em Solution method:}
The time-dependent GP equation is solved by the split-step Crank-Nicolson method by discretizing in space and time. The discretized equation is then solved by propagation, in either imaginary or real time, over small time steps. The contribution of the dipolar interaction is evaluated by a Fourier transformation to momentum space using a convolution theorem. The method yields the solution of stationary and/or non-stationary problems.

\noindent\\
{\em Reasons for the new version:}
Previously published dipolar Fortran and C programs \cite{dbec2015}, based on earlier programs and algorithms for GP equation with the contact interaction \cite{older}, are already used within the ultra-cold atoms community \cite{uca}. However, they are sequential, and thus did not allow for use of the maximum computing performance modern computers can offer. For this reason we have explored possible ways to accelerate our programs. Detailed profiling revealed that the calculation of FFTs is the most computationally demanding part of our programs. Since using GPUs to compute FFTs with optimized libraries like the cuFFT can lead to much better performance, we have decided to parallelize our programs using Nvidia CUDA toolkit. Also, the massive parallelism offered by GPUs could be exploited to parallelize the nested loops our programs have. We have focused on 2d and 3d versions of our programs, as they perform enough computation to justify and require the use of massive parallelism.

\noindent\\
{\em Summary of revisions:}
Previous C programs in two or three spatial dimensions are parallelized using CUDA toolkit from Nvidia and named similarly, with ``-cuda" suffix appended to their names. The structure of all programs is identical. Computationally most demanding functions performing time evolution (calcpsidd2, calcnu, calclux, calcluy, calcluz), normalization of the wave function (calcnorm), and calculation of physically relevant quantities (calcmuen, calcrms) were implemented as a series of CUDA kernels, which are executed on GPU. All kernels are implemented with grid-stride loops \cite{grid-stride}, which allow us to use the same kernel block sizes for all of our kernels. These block sizes can be changed in src/utils/cudautils.cuh, containing the optimal values for current Nvidia Tesla GPUs.

As before, CPU performs the initialization of variables and controls the flow of programs, offloading computation to GPU when needed. Because of the initialization, programs still require almost the same amount of CPU RAM as GPU RAM. Before any computation begins, relevant variables are copied to GPU, where they remain during computation, and only wave function array is returned back to CPU when it is required for writing output.

Parallelization with CUDA toolkit required some dynamically allocated arrays (tensors, matrices, or vectors) to become private for each GPU thread. This has caused an increase in the amount of used GPU memory, since the number of running threads on GPU is very large. Coupled with the fact that GPUs usually have smaller amount of RAM than CPU, this meant that our GPU versions of programs could be used for much smaller input in comparison to sequential versions. In order to fix this problem and reduce memory usage, our programs reuse temporary arrays as much as possible. Aside from allocation of complex tensor/matrix (for 3d or 2d case, respectively) in which we store wave function values, we allocate one complex tensor/matrix , and up to two double precision tensors/matrices, and reuse them for different purposes in computations. Allocated complex tensor/matrix is later also used as two double precision tensors/matrices, for other purposes. This required some reorganization of computation in several functions, mainly in calcmuen and calcpsidd2. In calcmuen we have reorganized computation to reuse temporary array and store partial derivatives in it, so instead of using three (in 3d) or two (in 2d) separate tensors/matrices for partial derivatives, we now use a single temporary tensor/matrix, which we also use for different purposes in other places in programs. In calcpsidd2 we have removed the use of additional temporary array that was only used in FFT computation, and also use real-to-complex and complex-to-real FFT transformations in place of complex-to-complex transformations of previous program versions. This change was possible because condensate density (input array for FFT) is purely real, and thus it exhibits Hermitian symmetry. Some FFT libraries, like the cuFFT used in these programs, can exploit this to reduce memory usage and provide better performance by calculating only non-redundant parts of the array. Additionally, programs can further reduce GPU RAM consumption by keeping the tensor/matrix used to store trap potential and dipolar potential in main RAM, configurable through POTMEM parameter in the input file. Setting value of POTMEM to 2 maximizes performance, and means that programs will allocate two separate tensors/matrices for storing trap potential and dipolar potential in GPU memory. This provides the best performance, but at the cost of a larger total memory consumption. If we set the value of POTMEM to 1, only one tensor/matrix will be allocated in GPU memory, to which trap potential and dipolar potential will be asynchronously copied from main memory when they are needed for computation. In this case, tensor/matrix will initially contain trap potential, which will be replaced with dipolar potential during execution of FFT in calcpsidd2, and replaced back with trap potential during inverse FFT. Finally, setting POTMEM to 0 will instruct the programs not to allocate any GPU memory for storing potentials and will instead use main memory, which GPU can access through slower PCI-Express bus. Figure~\ref{fig1} explains how memory is used and the possible values of POTMEM. We suggest using POTMEM value of 2 if memory permits, and using values of 1 or 0 if problem cannot fit into GPU memory. If POTMEM is not specified, programs will check if GPU memory is large enough to fit all variables and set POTMEM accordingly. 

\begin{figure}[!t]
\centering
\includegraphics[width=14cm]{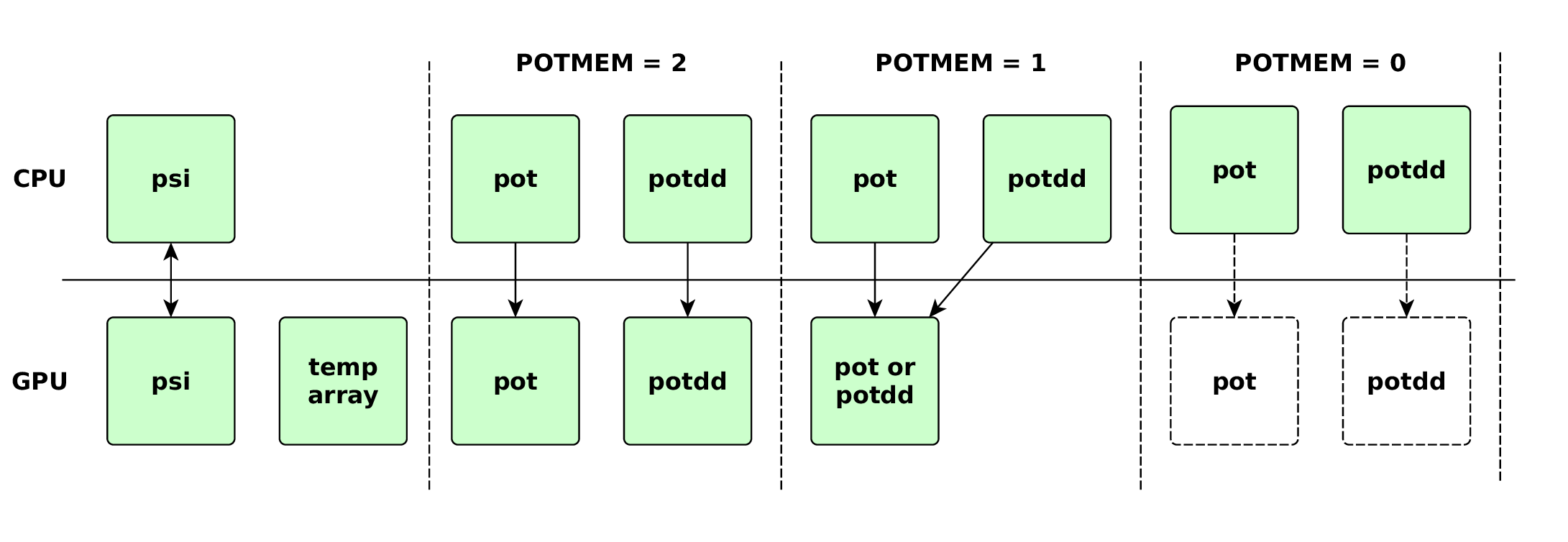}
\caption{Illustration of placement of relevant variables in CPU and GPU memory. CPU initializes its own wave function tensor/matrix (\textit{psi}), trap potential (\textit{pot}) and dipolar potential (\textit{potdd}), which is copied to GPU memory. Depending on value of POTMEM variable, GPU will either allocate the same tensors/matrices for trap and dipolar potential (POTMEM = 2), allocate only one tensor/matrix and use it for different purposes (POTMEM = 1), or will map \textit{pot} and \textit{potdd} from CPU and not allocate extra memory on GPU (POTMEM = 0). Additionaly, GPU allocates one complex tensor/matrix which is used for temporary data. This tensor/matrix is used either as a single complex tensor/matrix, or is divided into two double tensors/matrices which can then each contain the same number of elements as the complex tensor/matrix. }
\label{fig1}
\end{figure}

Time propagation functions calclux, calcluy, and calcluz have a recursive relation that makes them difficult to parallelize. In principle, recursive relations could be parallelized using a higher-order prefix sum algorithm \cite{prefix} (also known as scan algorithm), but implementation of this would require multiple CUDA kernels \cite{GPUprefix}. Since recursive relations are in the innermost loop, launching of all required kernels would create a sizeable overhead. Also, the number of grid points in each dimension is usually not large enough to compensate that overhead. Therefore, we have chosen an approach that, instead of parallelizing the inner loop which has the recursive relation, we parallelize the outer loops, and each GPU thread computes the whole innermost loop. Since each GPU thread now requires its own array for storing Crank-Nicolson coefficients cbeta, we reuse existing temporary tensor/matrix for storing these values. Similar pattern of parallelizing outer loops was also used in calcnorm, calcrms, and calcmuen.

We tested our programs at the PARADOX supercomputing facility at the Scientific Computing Laboratory of the Institute of Physics Belgrade. Nodes used for testing had Intel Xeon E5-2670 CPUs with 32 GB of RAM and Nvidia Tesla M2090 GPU with 6 GB of RAM. Figure~\ref{fig2} shows the speedup obtained for six DBEC-GP-CUDA programs compared to their previous versions \cite{dbec2015} executed on a single CPU core. Profiling reveals that the execution time is dominated by execution of FFTs and that the speedup varies significantly with changing of the grid size. This is due to FFT libraries used (FFTW in previous CPU version \cite{dbec2015} and cuFFT in this version), which use different algorithms for different input array sizes. We thus conclude that the best performance can be achieved by experimenting with different grid sizes around the desired target.

\begin{figure}[!t]
\centering
\includegraphics[width=7.3cm]{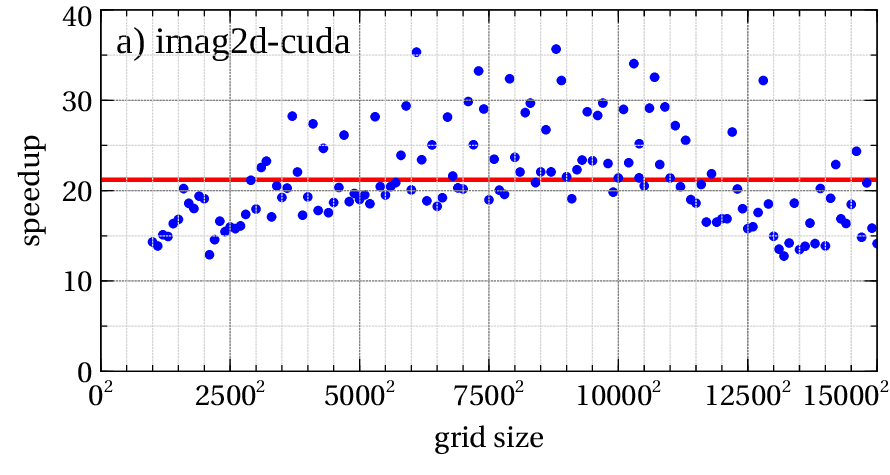}
\includegraphics[width=7.3cm]{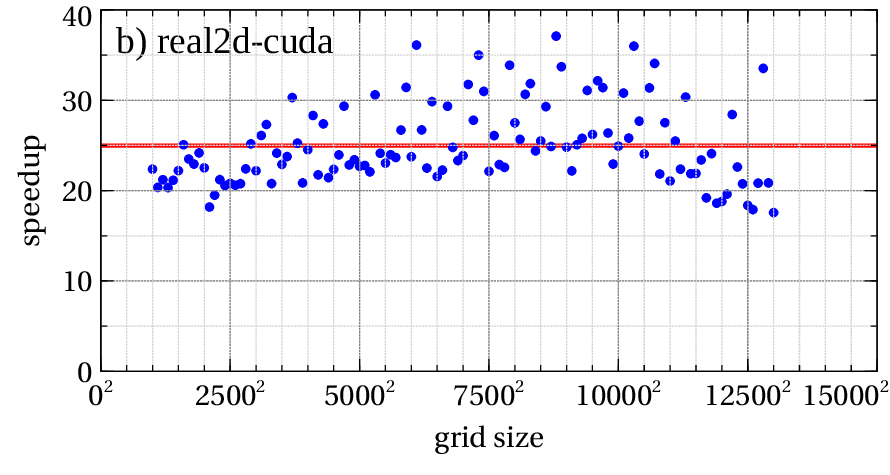}
\includegraphics[width=7.3cm]{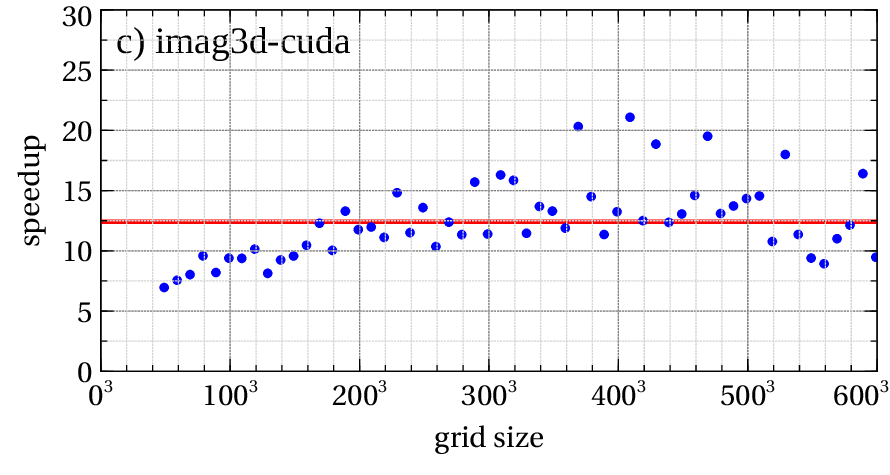}
\includegraphics[width=7.3cm]{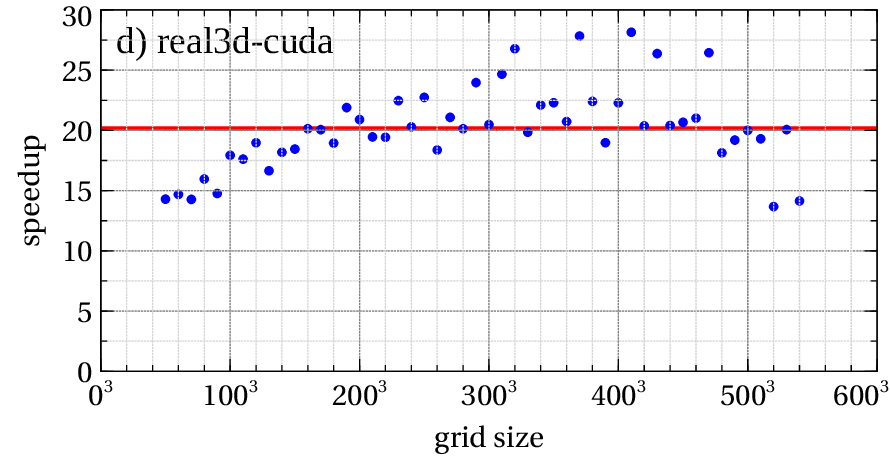}
\caption{Speedup in execution time of imag2dXY-cuda and imag2dXZ-cuda (top-left), real2dXY-cuda and real2dXZ-cuda (top-right), imag3d-cuda (bottom-left) and real3d-cuda (bottom-right) compared to the previous versions of programs \cite{dbec2015} executed on a single CPU core. Solid red line represents average speedup obtained. We tested linear grid sizes starting from $50^3$ in 3d and $1000^2$ in 2d, up to the maximum that could fit in GPU memory, which was $600^3$ for imag3d-cuda, $540^3$ for real3d-cuda, $15000^2$ for imag2dXY-cuda and imag2dXZ-cuda, and $13000^2$ for real2dXY-cuda and real2dXZ-cuda. Note that the dispersion of data is due to the use of FFTW\_ESTIMATE flag in library calls to FFTW in the CPU programs.}
\label{fig2}
\end{figure}

\noindent\\
{\em Restrictions:}\\
Programs will only run on computers with Nvidia GPU card (Tesla or GeForce) with Compute Capability 2.0 or higher (Fermi architecture and newer) and with CUDA toolkit installed (version 6.0 or higher).

\noindent\\
{\em Unusual features of all programs:}\\
As part of the memory usage optimizations, programs may slightly increase the number of spatial grid points in each dimension (NX, NY, NZ). This is due to FFT algorithms of cuFFT library that require additional memory to store temporary results. Our programs reuse already allocated memory to provide cuFFT with the temporary memory it requires, however, some problem sizes require much more memory, up to eight times more \cite{cuFFT}. For instance, if the number of grid points in any dimension is a large prime number, cuFFT uses an algorithm that requires eight times more memory than similarly sized power of two number. Adjustments of the number of grid points made in the programs ensure that cuFFT will not require such significantly increased additional memory. In case the programs perform the adjustments to grid size, this is reported in the output. 

\noindent\\
{\em Additional comments:}\\
This package consists of 6 programs, see Program title above. For the particular purpose of each program, please see descriptions below.

\noindent\\
{\em Running time:}\\
Example inputs provided with the programs take less than one minute on Nvidia Tesla M2090 GPU.

\noindent\\
Program summary (i)\\

\noindent
{\em Program title:} imag2dXY-cuda.\\
{\em Title of electronic files:} imag2dXY-cuda.cu and imag2dXY-cuda.cuh.\\
{\em Maximum RAM memory:} No upper bound.\\
{\em Programming language used:} CUDA C.\\
{\em Typical running time:} Minutes on a medium PC.\\
{\em Nature of physical problem:} This program is designed to solve the time-dependent dipolar nonlinear partial differential GP equation in two space dimensions in an anisotropic harmonic trap. The GP equation describes the properties of a dilute trapped Bose-Einstein condensate.\\
{\em Method of solution:} The time-dependent GP equation is solved by the split-step Crank-Nicolson method by discretizing in space and time. The discretized equation is then solved by propagation in imaginary time over small time steps. The method yields the solution of stationary problems.\\

\noindent\\
Program summary (ii)\\
\\
\noindent
{\em Program title:} imag2dXZ-cuda.\\
{\em Title of electronic files:} imag2dXZ-cuda.cu and imag2dXZ-cuda.cuh.\\
{\em Maximum RAM memory:} No upper bound.\\
{\em Programming language used:} CUDA C.\\
{\em Typical running time:} Minutes on a medium PC.\\
{\em Nature of physical problem:} This program is designed to solve the time-dependent dipolar nonlinear partial differential GP equation in two space dimensions in an anisotropic harmonic trap. The GP equation describes the properties of a dilute trapped Bose-Einstein condensate.\\
{\em Method of solution:} The time-dependent GP equation is solved by the split-step Crank-Nicolson method by discretizing in space and time. The discretized equation is then solved by propagation in imaginary time over small time steps. The method yields the solution of stationary problems.\\

\noindent Program summary (iii)\\
\\
\noindent
{\em Program title:} imag3d-cuda.\\
{\em Title of electronic files:} imag3d-cuda.cu and imag3d-cuda.cuh.\\
{\em Maximum RAM memory:} No upper bound.\\
{\em Programming language used:} CUDA C.\\
{\em Typical running time:} Tens of minutes on a medium PC.\\
{\em Nature of physical problem:} This program is designed to solve the time-dependent dipolar nonlinear partial differential GP equation in three space dimensions in an anisotropic harmonic trap. The GP equation describes the properties of a dilute trapped Bose-Einstein condensate.\\
{\em Method of solution:} The time-dependent GP equation is solved by the split-step Crank-Nicolson method by discretizing in space and time. The discretized equation is then solved by propagation in imaginary time over small time steps. The method yields the solution of stationary problems.\\

\noindent Program summary (iv)\\
\\
\noindent
{\em Program title:} real2dXY-cuda.\\
{\em Title of electronic files:} real2dXY-cuda.cu and real2dXY-cuda.cuh.\\
{\em Maximum RAM memory:} No upper bound.\\
{\em Programming language used:} CUDA C.\\
{\em Typical running time:} Tens of minutes on a good workstation.\\
{\em Unusual feature:} If NSTP=0, the program requires and reads the file imag2dXY-den.txt, generated by executing imag2dXY-cuda with the same grid size parameters.\\
{\em Nature of physical problem:} This program is designed to solve the time-dependent dipolar nonlinear partial differential GP equation in two space dimensions in an anisotropic harmonic trap. The GP equation describes the properties of a dilute trapped Bose-Einstein condensate.\\
{\em Method of solution:} The time-dependent GP equation is solved by the split-step Crank-Nicolson method by discretizing in space and time. The discretized equation is then solved by propagation in real time over small time steps. The method yields the solution of dynamical problems.\\

\noindent Program summary (v)\\
\\
\noindent
{\em Program title:} real2dXZ-cuda.\\
{\em Title of electronic files:} real2dXZ-cuda.cu and real2dXZ-cuda.cuh.\\
{\em Maximum RAM memory:} No upper bound.\\
{\em Programming language used:} CUDA C.\\
{\em Typical running time:} Tens of minutes on a good workstation.\\
{\em Unusual feature:} If NSTP=0, the program requires and reads the file imag2dXZ-den.txt, generated by executing imag2dXZ-cuda with the same grid size parameters.\\
{\em Nature of physical problem:} This program is designed to solve the time-dependent dipolar nonlinear partial differential GP equation in two space dimensions in an anisotropic harmonic trap. The GP equation describes the properties of a dilute trapped Bose-Einstein condensate.\\
{\em Method of solution:} The time-dependent GP equation is solved by the split-step Crank-Nicolson method by discretizing in space and time. The discretized equation is then solved by propagation in real time over small time steps. The method yields the solution of dynamical problems.\\

\noindent Program summary (vi)\\
\\
\noindent
{\em Program title:} real3d-cuda.\\
{\em Title of electronic files:} real3d-cuda.cu and real3d-cuda.cuh.\\
{\em Maximum RAM memory:} No upper bound.\\
{\em Programming language used:} CUDA C.\\
{\em Typical running time:} Tens of minutes on a good workstation.\\
{\em Unusual feature:} If NSTP=0, the program requires and reads the file imag3d-den.txt, generated by executing imag3d-cuda with the same grid size parameters.\\
{\em Nature of physical problem:} This program is designed to solve the time-dependent dipolar nonlinear partial differential GP equation in three space dimensions in an anisotropic harmonic trap. The GP equation describes the properties of a dilute trapped Bose-Einstein condensate.\\
{\em Method of solution:} The time-dependent GP equation is solved by the split-step Crank-Nicolson method by discretizing in space and time. The discretized equation is then solved by propagation in real time over small time steps. The method yields the solution of dynamical problems.\\

\section*{Acknowledgements}
\noindent
 V.~L., A.~B., A.~B., and S.~\v S acknowledge support by the 
Ministry of Education, Science, and Technological Development of the Republic of Serbia under 
projects ON171017, III43007, ON174023, and IBEC, and by the DAAD - German Academic and 
Exchange Service under project IBEC.
P.~M. acknowledges support by the Science and Engineering Research Board, Department of Science and Technology, Government of India under project No.~EMR/2014/000644.
S.~K.~A. acknowledges support by the CNPq of Brazil under project 303280/2014-0, and by the FAPESP of Brazil under project 2012/00451-0.
Numerical simulations were run on the PARADOX supercomputing facility
at the Scientific Computing Laboratory of the Institute of Physics
Belgrade, supported in part by the Ministry of Education, Science,
and Technological Development of the Republic of Serbia under project
ON171017.

\end{small}

\end{document}